\documentclass[oldversion]{aa} 
\usepackage{graphicx}
\usepackage{longtable}
\usepackage{txfonts}
\usepackage{rotating}
\usepackage{lscape}
\newcommand{\gsim}{\;\lower.6ex\hbox{$\sim$}\kern-7.75pt\raise.65ex\hbox{$>$}\;}
\newcommand{\lsim}{\;\lower.6ex\hbox{$\sim$}\kern-7.75pt\raise.65ex\hbox{$<$}\;}

\begin{document}
\title{Multiple populations in $\omega$~Centauri: a cluster analysis of spectroscopic data
 }

\author{
R.G. Gratton\inst{1},
C.I. Johnson\inst{2}\inst{3},
S. Lucatello\inst{1},
V. D'Orazi\inst{1},
\and
C. Pilachowski\inst{4}
}

\authorrunning{R.G. Gratton}
\titlerunning{Multiple populations in $\omega$~Cen}

\offprints{R.G. Gratton, raffaele.gratton@oapd.inaf.it}

\institute{
INAF-Osservatorio Astronomico di Padova, Vicolo dell'Osservatorio 5, I-35122
 Padova, Italy
\and
Department of Physics and Astronomy, University of California, Los Angeles, 430 Portola Plaza, Box
951547, Los Angeles, CA 90095-1547, USA; cijohnson@astro.ucla.edu;
\and
National Science Foundation Astronomy and Astrophysics Postdoctoral Fellow
\and
Department of Astronomy, Indiana University, Swain West 319, 727 East Third Street, Bloomington,
IN 47405-7105, USA; catyp@astro.indiana.edu  }

\date{}
\abstract{$\omega$~Centauri, the largest globular cluster of the Milky Way, is 
composed of several stellar populations, which may be evidenced from both 
photometry and spectroscopy. The history of how these different populations 
assembled might allow us to reconstruct the evolution of this complex object. 
In particular, understanding the detailed chemical evolution would be particularly
illuminating.  However, this is not easy because of the errors intrinsic
to abundance determinations. We performed 
a statistical cluster analysis on the large data set of accurate abundances 
recently provided by Johnson and Pilachowski (2010) for about 800 red giant 
branch stars. We find that stars in $\omega$~Cen divide into three main groups. 
The metal-poor group includes about a third of the total. It shows a moderate 
O-Na anticorrelation, and similarly to other clusters, the O-poor second generation 
stars are more centrally concentrated than the O-rich first generation ones. 
This whole population is La-poor, with a pattern of abundances for $n-$capture 
elements which is very close to a scaled r-process one. The metal-intermediate 
group includes the majority of the cluster stars. This is a much more complex 
population, with an internal spread in the abundances of most elements. It shows 
an extreme O-Na anticorrelation, with a very numerous population of extremely 
O-poor and He-rich second generation stars. This second generation is very 
centrally concentrated. This whole population is La-rich, with a pattern of the 
abundances of $n-$capture elements that shows a strong contribution by the $s-$process.
The spread in metallicity within this metal-intermediate population is not very 
large, and we might attribute it either to non uniformities of an originally very 
extended star forming region, or to some ability to retain a fraction of the 
ejecta of the core collapse SNe that exploded first, or both. As previously
noticed, the metal-rich group has an Na-O correlation, rather than anticorrelation. 
There is evidence for the contribution of both massive stars ending their life as 
core-collapse SNe, and intermediate/low mass stars, producing the s-capture 
elements. Kinematics of this population suggests that it formed within the cluster
rather than being accreted. }

\maketitle

\section{Introduction}

It has recently become evident that globular clusters, hitherto considered as simple 
stellar populations, are actually made of multiple stellar populations
(see Gratton et al. 2001, 2004). Evidence includes both photometry (Bedin et al.
2004; Piotto et al. 2007) and spectroscopy (Gratton et al. 2001; Carretta et al. 2009a, 
2009b, 2010a). Most globular clusters host only a small number of stellar populations, 
differing in their content of light elements, typically described by anticorrelations
among C and N, Na and O, Mg and Al, and likely He (Bedin et al. 2004; Carretta et al. 
2009a; Gratton et al. 2010), while the abundances of Fe-peak elements do not show any 
spread (Carretta et al. 2009c). However in a few, generally massive globular clusters, 
different populations differ in the abundances of virtually all elements. On many respect, 
these objects can be considered as intermediate between globular clusters and ultra compact 
dwarf galaxies (see e.g. Forbes \& Kroupa 2011 and Norris \& Kannappan 2011). Examples 
include M54 (Carretta et al. 2010b), M22 (Marino et al. 2009), NGC1851 (Lee et al. 2009;
Carretta et al. 2010d) and $\omega$~Cen. This last, which is the brightest and most 
massive Milky Way cluster, represents the most extreme case of such variations.

The presence of multiple populations in $\omega$~Centauri was discovered almost half
a century ago by Woolley (1966). Various studies provided evidence for the large spread in 
metallicity within this cluster: among many others, we may cite Freeman \& Rodgers (1975), 
Butler et al. (1978), Cohen (1981), Norris \& Da Costa (1995a, 1995b), Suntzeff \& 
Kraft (1996), and Smith et al. (2000). All these studies found a
predominance of metal-poor stars ([Fe/H]$<-1.5$), with a tail up to rather high 
metallicities ([Fe/H]$\sim -1$). More recently, Pancino et al. (2000) discussed the 
presence of a group of metal-rich stars on the red of the main red giant branch (RGB), 
which they called RGB-a (for anomalous), and have a metallicity [Fe/H]$>-1$. Perhaps 
the most surprising discovery was however the splitting of the main sequence (Bedin et 
al. 2004) into at least two (and possibly more) separate sequences, and the fact
that the red sequence is clearly more numerous than the blue one. This lead to the
suspicion that the blue sequence is much more He-rich (Norris 2004), a fact soon
confirmed by the spectroscopic analysis by Piotto et al. (2005) showing that the
blue main sequence is more metal-rich than the red one, just the opposite of what
should be expected if the splitting were to be attributed simply to a metal abundance
difference.

Many different populations are clearly present in $\omega$~Cen: multiple
populations are found in the RGB (Sollima et al. 2005a), subgiant
branch and main sequence (Bellini et al. 2010). Early attempts to reconstruct 
the history of these populations met severe problems. For instance, various authors 
proposed age-metallicity relations for $\omega$~Cen by combining photometric and 
spectroscopic observations of subgiant branch stars (see e.g. Sollima et al. 2005b; 
Stanford et al. 2006; Villanova et al. 2007); however results were contradictory with each other,
likely because these studies neglected the large variations in He content that are
present among different groups of stars. However, reconstructing early history of
globular clusters from photometry alone is very difficult, because the long time elapsed
since the cluster formation makes differences due to ages subtle, and easily masked
by other effects (He and heavy element variations). Some clarification might then possibly 
come from the chemistry, exploiting the fact that different elements are produced by stars in
different mass ranges, and hence on different timescales. Thanks to the use of multi-fibre
instruments, very extensive high resolution spectroscopic studies of several hundred red giants are now 
available (Johnson \& Pilachowski 2010; Marino et al. 2011). These studies provide several
interesting observations. For instance, while a clear Na-O anticorrelation is present among 
metal-poor stars, overabundances of both Na and O are obtained for the most metal-rich ones 
(those on the RGB-a of Pancino et al. 2000). In addition, the Na-O anticorrelation is present
in stars over a large range in metallicity, possibly its extension increasing with
metallicity. This is not easy to be reconciled with the typically very narrow metal 
distribution of other clusters. Finally, the abundances of the $n-$capture elements 
mainly produced by the $s-$process clearly rise with metallicity (Norris \& Da Costa,
1995; Smith et al. 2000; Pancino 2003; Johnson \& Pilachowski 2010; Marino et al. 2011).
The production of s-process elements is usually thought 
to occur in small mass asymptotic giant branch (AGB) stars, and on a much longer timescale than that considered 
for the evolution of the stars responsible of the Na-O anticorrelation from fast
rotating massive stars (Decressin et al. 2007) or massive AGB stars undergoing hot bottom
burning (Ventura et al. 2001). As discussed by Marino et al. (2011), there is a timescale
problem to be solved.
 
In addition, Carretta et al. (2010a) proposed that $\omega$~Cen
might be similar to M~54, a GC whose position is coincident with the nucleus of the
Sagittarius dwarf spheroidal (dSph) galaxy. They noticed that when this galaxy will be dispersed
as a consequence of its tidal interaction with the Milky Way, some part of the stars
in the field of the dSph - themselves much more metal-rich than M54 - would likely remain 
locked within M~54. The red giants would then appear as an anomalous red giant branch, in analogy with the
RGB-a of $\omega$~Cen. Whether this is the correct explanation of the RGB-a of $\omega$~Cen
remains to be verified. This might possibly be obtained by a chemical abundance analysis
of the RGB-a stars. For instance, if this interpretation were correct, we should expect
that the RGB-a stars be poor in $\alpha$-elements, since this is the typical signature
observed in metal-rich stars in dSph.

Part of the analysis problems likely arise from the difficulties of clearly separating the
different populations of $\omega$~Cen. While progress is being made to understand the separation
of the different populations on the main sequence, the separation of the populations on
the RGB remains difficult. Detailed abundances for many elements can help to differentiate
the RGB populations and understand their origins.
To explain their results, Marino et al. (2011) proposed a 
subdivision of their RGB sample into different metallicity bins, but realized that
some stars could be assigned to incorrect bins due to errors in metallicity. They
attempted to correct for such errors, but their corrections were not based on robust,
objective criteria. In this paper, we 
wish to re-examine this issue using a more objective approach, that is using grouping algorithms 
usually adopted in statistical analysis. These methods are applied to the sample by Johnson \& 
Pilachowski (2010), which is about three times more numerous than that considered by Marino et 
al. (2011), while with similar error bars. We will show that a few groups naturally arise from 
the data itself. These groups, derived from chemical properties, also have different broad band colours. 
The properties of these groups nicely correspond to the main populations found in other
evolutionary phases (e.g. the main sequence). We also briefly discuss dynamical properties
for these groups. We think that such a division in groups will allow significant
progress in future modelling of the formation of this cluster.

\section{Input data}

The data we considered in our analysis are those produced by Johnson \& Pilachowski (2010),
who provided abundances for eleven elements (Fe, O, Na, Al, Si, Ca, Sc, Ti, Ni, La, and Eu)
in 855 stars. In addition to the elemental abundances, we have coordinates (and hence
distance from cluster center), radial velocities and photometry (BV from van Loon
et al. 2007, and JHK from 2MASS, Skrutskie et al 2006). 

The main objective of our analysis is to find ``natural groups" from this data set,
that can help in the interpretation of the evolution of $\omega$~Cen. Group analysis
usually requires several parameters to be known for all the members of the input
populations. Due to availability of data, not all elements were 
measured in all stars, a fact that should be taken into account in our analysis.
Furthermore, while error bars for individual elements can be significant with
respect to the overall scatter throughout the cluster, it is also clear that there 
is some redundancy among the input data, since various elements can be grouped 
together having similar nucleosynthesis. For instance, abundances of the Fe-peak
elements Sc and Ni (not available for many stars) do not really add much to what 
is obtained using Fe alone. We then considered in our analysis the following
four quantities:
\begin{itemize}
\item The Fe abundance [Fe/H], which is assumed to be representative of the overall 
metallicity. We will see later that this quantity is well correlated with the colour 
of stars along the RGB, as expected. Fe may be produced by both core collapse and 
thermonuclear supernovae.
\item The ratio between the Na and O abundances [Na/O]. This ratio is representative of
the location of stars along the Na-O anticorrelation, and is likely correlated with
the He abundance, although this relation is presumably not linear (see e.g. Gratton
et al. 2010)
\item The average of the elements Si, Ca, and Ti, all mainly produced by 
$\alpha-$capture reactions in massive stars, later exploding as core-collapse SNe.
Use of an average of these three elements has two advantages: first, reduces the
number of stars for which data are not available; second, it reduces star-to-star
scatter. In practice, we used this [$\alpha$/Fe] ratio. This might be either a
measure of the relative contribution of core collapse and thermonuclear SNe, or
(more likely in the case of $\omega$~Cen, see later) of the weights to be given to
the contribution by stars of different masses among the core collapse SNe.
\item The abundance of La, which is an $n-$capture element that in the Sun is mainly
produced by the $s-$process.
\end{itemize}

These data are available for 797 stars, that is more than 93\% of the total
sample observed by Johnson \& Pilachowski (2010). A cluster analysis for this
sample is then expected to be representative at least for the major populations in
$\omega$~Cen. Most of the 70 stars excluded in this analysis lack La abundances (57
stars); the remaining ones lack either O or Na abundances (or both). Missing
stars tend to be located farther from the cluster center (average distance of
$7.8\pm 0.5$~arcmin, with respect to an average value of $6.30\pm 0.16$~arcmin 
for the stars included in the sample). Likely for this reason, they have on average a 
larger [O/Fe] value ($0.37\pm 0.02$\ vs $0.12\pm 0.01$) (see Section 5.2). However,
their Na abundance is not smaller than for the stars of our sample ($0.25\pm 0.04$\ vs $0.16\pm 0.01$),
and the Na-O anticorrelation is poorly defined for these stars - likely because
of larger than average errors. The excluded stars have on 
average an [Fe/H] slightly lower than for those stars included in the sample 
([Fe/H]=$-1.69\pm 0.03$\ vs $-1.60\pm 0.01$). It is then likely that most of
the stars excluded belong to groups \#4 an \#5 cited below, although they are likely
distributed among all groups. Three of the excluded stars (45485, 4715, and especially 45358)
have large La, as well as Na and Al abundances. No O abundance determination is
available for these stars, all of them being metal rich ([Fe/H]$>-1.1$)

Of course, it would be important to use additional data in our classification, but unluckily 
these are available
only for a minority of the stars. For instance, Al abundances help to better clarify
the p-capture mechanisms responsible for the Na-O anticorrelation. Al abundances
have been estimated by Johnson et al. (2008, 2009, 2010) for 311 stars of our sample. In addition 
Eu abundances would be very helpful to better understand the nature of the $n-$capture 
processes that are relevant for $\omega$~Cen. However, Eu abundances are available for 
only 184 of these 797 stars. We deem reduction of our sample to only those stars having 
also Al or Eu abundances too much a limitation. On the other hand, in Sections 3.3 and 3.4 we will 
discuss both Al and Eu abundances for the groups we will derive from our analysis. 

\section{Analysis and results}

In our analysis we used the $R$\ statistical package (R Development Core Team, 2011). $R$\ is a system 
for statistical computation and graphics, freely available 
on-line\footnote{http://www.R-project.org}. Several algorithms for cluster analysis
are available within $R$. Our results are based on the $k-$means algorithm 
(Steinhaus 1956; MacQueen 1967), although in Section 4 we will briefly discuss
results obtained with other clustering algorithms. $k-$means clustering is a method of cluster analysis 
which aims to partition $n$\ observations into $k$\ clusters in which each observation belongs 
to the cluster with the nearest mean. Given a set of observations ($x_1$, $x_2$, ..., 
$x_n$), where each observation is a $d$-dimensional real vector, $k-$means clustering 
distribute the $n$\ observations into $k$\ sets ($k=n$) $S=\{S_1, S_2, ..., S_k\}$\
so as to minimize the within-cluster sum of squares $W$:
\begin{equation}
W = arg min \sum\limits_{i=1}^k \sum\limits_{x_j \in S_i} \| x_j - \mu_i \|^2
\end{equation}
Practically, there are various algorithms used to find this minimum. Within the R-package, 
the algorithm of Hartigan and Wong (1979) is used by default. The Hartigan and Wong algorithm 
generally does a better job than other $k-$means algorithms. 

Two key features of $k-$means are:
\begin{itemize}
\item Euclidean distance is used as a metric and variance is used as a measure of 
cluster scatter. The various parameters used might be transformed before computing
these distances, in order to weight them adequately. A sensible choice is to normalize
them to the observational errors. These are quite similar for the different quantities
we considered (see Johnson \& Pilachowski, 2010), so that we skipped this step in our
analysis.
\item The number of clusters $k$\ is an input parameter: an inappropriate choice of $k$\ 
may yield poor results. That is why, when performing $k-$means, it is important to run 
diagnostic checks for determining the number of clusters in the data set.
\end{itemize}
A key limitation of $k-$means is its cluster model. The concept is based on spherical
clusters that are separable in a way so that the mean value converges towards the cluster 
center. The clusters are expected to be of similar population, so that the assignment to 
the nearest cluster center is the correct assignment. When the clusters have very
different size, this may result in poor assignation of members to clusters. However,
in the present case, we expect that with appropriate choice of the number of
groups $k$, most of the scatter within one cluster is due to
observational errors, which are similar for the different clusters. We then 
considered the assumption of similar size for the different clusters acceptable.
Of course, this does not mean that occasionally assignation of some member (that
is star) to a particular cluster is questionable.

In general, there is a strong inter-relation between results of cluster analysis
provided by $k-$means and those provided by the principal component analysis. It is 
then useful to review the role played by the different parameters considered in our 
analysis in this perspective. On the whole, [$\alpha$/Fe], which has a very small 
intrinsic scatter, comparable to observational errors, plays a minor role, and its 
inclusion in the analysis does not change group identification significantly. The 
other three parameters considered in Section 2 are much more important, but since 
there is a strong positive 
correlation between [Fe/H] and [La/Fe], the plane of the two principal components is 
very close to the [Na/O] vs [La/H] plane. Then subdivision of stars in different 
clusters may be well visualized in this plane, and we will use these two quantities
as proxy of the first two principal components.

\begin{center}
\begin{figure}
\includegraphics[width=8.8cm]{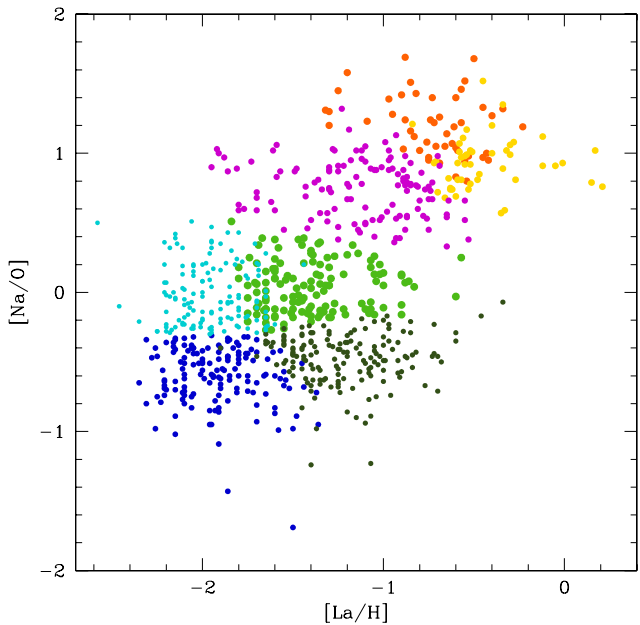}
\caption{Assignation of stars to the different groups in our k-means analysis in 
the [La/H] vs [Na/O] plane. This plane roughly corresponds to that of the second 
and first component in a principal component analysis. Different colours are for 
stars of the different groups: green: group \#1; red: group \#2a; yellow: group \#2b; 
violet: group \#3; blue: group \#4; black : group \#5; cyan: group \#6.}
\label{f:fig1}
\end{figure}
\end{center}

As mentioned above, a critical assumption in the $k-$means analysis is the 
number of clusters. In our analysis, we started considering a small number of 
clusters ($k=3$). In this case, we found assignation of stars to the different
clusters to be only driven by their [Na/O] value, which is the quantity exhibiting
the largest scatter, and as mentioned above is close to the first principal component
in our data. This result is of limited interest. We then increased
$k$, looking for a combination that would distinguish between
groups having different metallicity (actually [La/H], roughly the second principal component), 
because we expect this to be a major parameter driving distinction in several 
populations within $\omega$~Cen.
Since the run of [Na/O] is very extended among metal-rich stars, and guided by the
result we obtained with $k=3$, we deemed that
three groups might be required for this metal abundance range, while perhaps two
groups might be enough for the more metal-poor stars, where the [Na/O] values are 
apparently less scattered. Indeed, an analysis with $k=5$\ separates stars
among both first and second principal components. However, the separation along
the second component appeared much cleaner with 6 groups (see Figure~\ref{f:fig1}).

\begin{table*}[htb]
\centering
\caption[]{Main group parameters}
\begin{scriptsize}
\begin{tabular}{cccccccccccccc}
\hline
Group&Colour&Datum &N. stars&[m/H]	&[Fe/H]	&[$\alpha$/Fe]&[O/Na]&	[O/Fe]&	[Na/Fe]	&[Al/Fe]&[La/Fe]&[Eu/Fe]&RV \\
     &code& & &	&	& & & &	& & & &(km/s)\\
\hline
\multicolumn{14}{c}{Metal-Poor groups}\\
\hline
\#4 &Blue     &avg	&148 &-1.738&-1.758	&0.20	& 0.60	& 0.39	&-0.22	& 0.34 &-0.16	&0.16 & $232.4\pm 1.1$\\
	&       &st.dev.&    & 0.118& 0.133	&0.08	& 0.21	& 0.11	& 0.20	& 0.24 & 0.17	&0.20 & 14.5 \\
\#6 &Turquoise&avg	&132 &-1.752&-1.794	&0.21	& 0.01	& 0.18	& 0.17	& 0.62 &-0.17	&0.20 & $231.6\pm 1.2$\\
	&       &st.dev.&    &0.115	&0.117	&0.08	& 0.22	& 0.18	& 0.17	& 0.30 & 0.18	&0.25 & 13.9 \\
\hline
\multicolumn{14}{c}{Metal-Intermediate groups}\\
\hline
\#5 &Dark green&avg	&162 &-1.562&-1.575	&0.27	& 0.47	& 0.46	&-0.01	& 0.34 & 0.35	&0.11 & $232.9\pm 1.0$\\
	&       &st.dev.&    & 0.182& 0.196	&0.12	& 0.18	& 0.13	& 0.16	& 0.23 & 0.16	&0.21 & 13.1 \\
\#1 &Bright green&avg	&132 &-1.642&-1.696	&0.28	&-0.04	& 0.23	& 0.27	& 0.69 & 0.30	&0.18 & $230.8\pm 1.3$\\
	&       &st.dev.&    & 0.138& 0.172	&0.11	& 0.18	& 0.18	& 0.17	& 0.32 & 0.18	&0.28 & 14.5 \\
\#3 &Magenta&avg	&127 &-1.600&-1.523	&0.32	&-0.74	&-0.38	& 0.37	& 1.04 & 0.39	&0.18 & $230.6\pm 1.3$\\
	&       &st.dev.&    & 0.120& 0.174	&0.08	& 0.20	& 0.23	& 0.14	& 0.21 & 0.26	&0.16 & 14.7 \\
\#2a &Red   &avg	& 49 &-1.444&-1.308	&0.36	&-1.20	&-0.66	& 0.54	& 0.98 & 0.58	&0.17 & $232.3\pm 2.2$\\
	&       &st.dev.&    & 0.177& 0.156	&0.08	& 0.22	& 0.26	& 0.14	& 0.15 & 0.20	&0.18 & 15.3 \\
\hline
\multicolumn{14}{c}{Metal-Rich group}\\
\hline
\#2b &Yellow&avg	& 47 &-1.176&-0.920	&0.36	&-0.93	&-0.19	& 0.74	& 0.67 & 0.49	&0.20 & $230.8\pm 1.3$\\
	&      &st.dev.&    & 0.212& 0.192	&0.14	& 0.19	& 0.26	& 0.26	& 0.30 & 0.18	&0.26 &  9.1 \\
\hline
\end{tabular}
\end{scriptsize}
\label{t:tab1}
\end{table*}

As we will see in the discussion, there are many appealing features in this subdivision; 
however, a first exploration
of the results shows that one of the clusters (\#2) is likely the results of the
combination of two groups of stars, with quite different characteristics. We therefore
performed a $k-$means analysis of this group alone, and divided it into two further
groups, which we called \#2a and \#2b. Thus, at the end, our analysis is based on 7 
groups\footnote{It should be noticed that if we had adopted 7 groups at start of
our analysis, the subdivision would have been almost identical to those obtained
with the 6 groups analysis, but the most metal and La-poor stars would have
been divided into three groups with decreasing [O/Na] values, rather than two.
This is different from the subdivision of group \#2, adopted throughout
this paper. We think that our approach provides a better insight into the properties
of $\omega$~Cen. However, it is clear that there is some arbitrariness in the way 
we performed our cluster analysis.}. The main characteristics of the different 
groups are summarized in
Table~\ref{t:tab1}. Briefly:
\begin{itemize}
\item Two groups are made of metal-poor stars, either O-rich (group \#4, blue 
in all figures) or O-intermediate (group \#6, turquoise). The [Fe/H] value is 
very similar, and all these stars 
are La-poor. Together, these two groups have 280 stars, that is 35\% of the
sample.
\item Four groups are made of stars of intermediate metallicity. They are, in
order of increasing [Na/O]: group \#5 (dark green), \#1 (bright green), 
\#3 (magenta), and \#2a (red). This order also
roughly corresponds to increasing [Fe/H] values (with an inversion between the
two first groups). All these stars are La-rich. In total, these groups have 
470 stars, that is 59\% of the sample.
\item Finally, one group (\#2b, yellow) is made of metal-rich stars. These are
both Na and O-rich, and also La-rich. This group is made of 47 stars, that is 6\%
of the sample.
\end{itemize}
The properties of these different groups will be examined in more detail in the
rest of this section.

\begin{center}
\begin{figure}
\includegraphics[width=8.8cm]{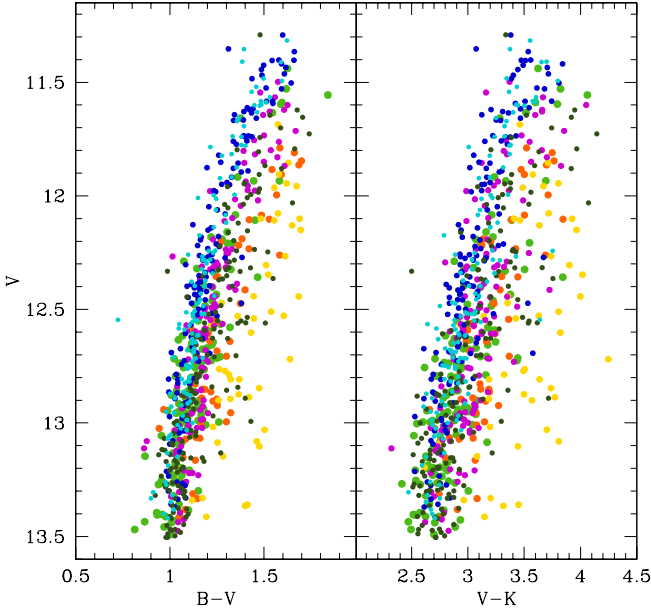}
\caption{V-(B-V) (Upper panel) and V-(V-K) (lower panel) colour-magnitude diagrams for the observed stars in $\omega$~Cen. Different colours are for stars of the different groups (see
Figure~\ref{f:fig1}). }
\label{f:fig2}
\end{figure}
\end{center}

\begin{center}
\begin{figure}
\includegraphics[width=8.8cm]{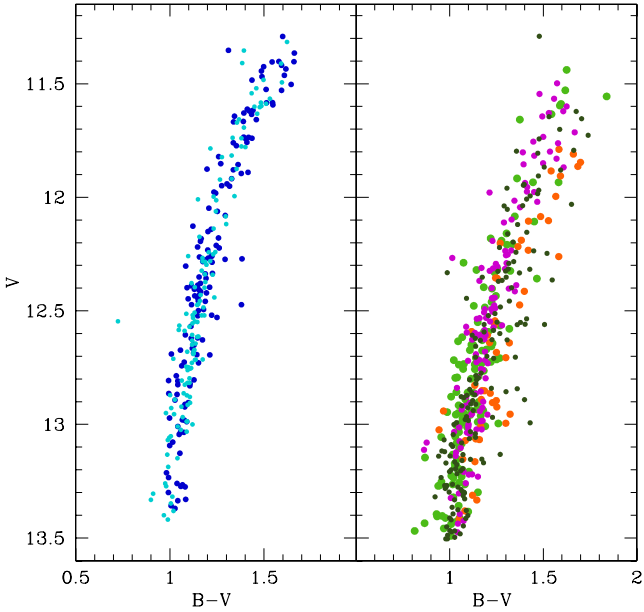}
\caption{V-(B-V) colour-magnitude diagram for the observed
stars in $\omega$~Cen. Metal poor groups (\#4 and \#6) are in the left panel, metal
intermediate ones (\#1, \#2a, \#3, \#5) in the right one. Different colours are for stars 
of the different groups (see Figure~\ref{f:fig1}).}
\label{f:fig3}
\end{figure}
\end{center}

\subsection{Colour-magnitude diagram}

Figure~\ref{f:fig2} shows the (V, B-V) and the (V, V-K) colour-magnitude diagrams for
the stars observed in $\omega$~Cen. Stars assigned to the different groups
are plotted with different colours. From this diagram, it is clear that the
stars assigned to the metal-poor groups (\#4 and \#6, blue and cyan respectively) are on the blue side
of the RGB, while those assigned to the metal-rich group (\#2b, yellow) are
on the red side. This last group can then be readily identified with the
RGB-a sequence of Pancino et al. (2000). The metal-poor groups define a
tight sequence, with no obvious segregation between the two groups, which
indeed have very similar mean [Fe/H] and only differs for their [Na/O] ratio. 
This is most clearly shown in Figure~\ref{f:fig3}, where we plotted separately the (V, B-V) 
diagrams for the metal-poor groups (left panel),
and for those of intermediate metallicity (right panel).

From these figures the presence of a correlation between colours along the
RGB and stellar metallicity is apparent, as expected from evolutionary models.
We may quantify this dependence, to
produce a metallicity index which depends on the colours of the stars along
the RGB. The usefulness of this index will be discussed later. In order
to define it, we started by fitting a cubic polynomial through the RGB of
the two metal-poor groups. We then measured the (B-V) offset $\Delta$(B-V) between each
star and this reference sequence. Finally, we fit a bilinear relation
between [Fe/H] (dependent variable), and $\Delta$(B-V) and V (independent variables).
The best fit parameters from this relation allows to have for each star a 
parameter, that we call [m/H], which on average is identical to [Fe/H], but
that obviously can be different from it for each individual star (and even
for each group). The average value of [m/H] for each group is also listed in
Table~\ref{t:tab1}.

\begin{center}
\begin{figure}
\includegraphics[width=8.8cm]{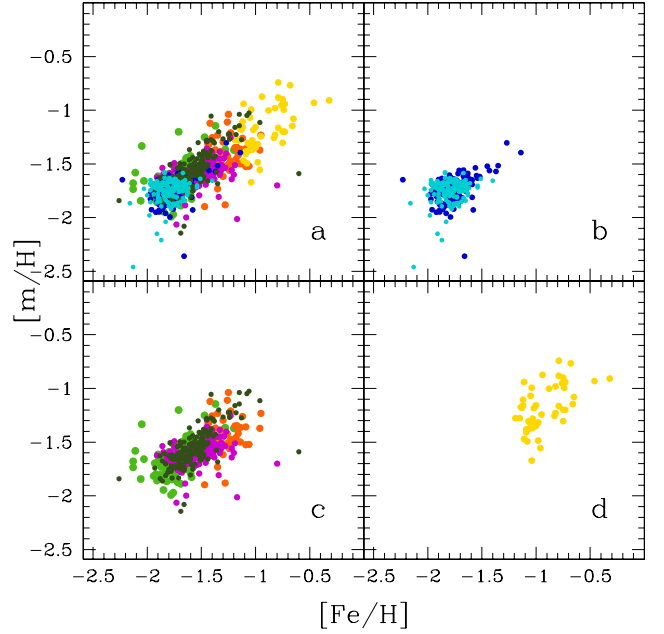}
\caption{Comparison between Iron abundances from spectroscopy [Fe/H] and
metal abundances derived from colours along the RGB ([m/H]). Different 
colours are for stars  of the different groups (see Figure~\ref{f:fig1}).
Results for all stars are plotted in panel a; panel b is is only for
metal-poor stars; panel c is only for metal-intermediate stars; and panel d
is only for metal-rich stars.}
\label{f:fig4}
\end{figure}
\end{center}

\begin{center}
\begin{figure}
\includegraphics[width=8.8cm]{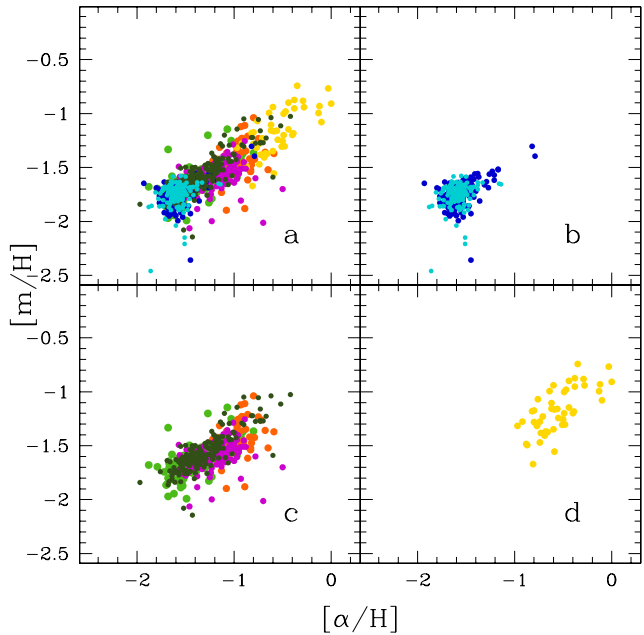}
\caption{[$\alpha$/H] vs
metal abundances derived from colours along the RGB ([m/H]). Different 
colours are for stars  of the different groups (see Figure~\ref{f:fig1}).
Results for all stars are plotted in panel a; panel b is is only for
metal-poor stars; panel c is only for metal-intermediate stars; and panel d
is only for metal-rich stars.}
\label{f:fig4a}
\end{figure}
\end{center}

Figure~\ref{f:fig4} shows the correlation between [m/H] and [Fe/H] for the whole sample,
as well as separately for the metal-poor, metal-intermediate, and metal-rich
groups. Of course, there is a good overall agreement between [m/H] and [Fe/H], the
linear correlation coefficient being r=0.58, which significance is extremely
high given the size of the sample (797 stars). An even better correlation 
is obtained with [$\alpha$/H] (r=0.64; see Fig~\ref{f:fig4a}), 
while the correlations are markedly poorer with e.g. [Na/H] or [O/H]. This
confirms that [m/H] is a good proxy for a combination of
[Fe/H] and [$\alpha$/H] (these two quantities being extremely well
correlated each other, with r=0.90).

We notice that when examining the results for the individual groups, while
there is still a good correlation between [m/H] and [Fe/H] for the groups
of intermediate metallicity, this correlation is much less obvious for the
other groups. In particular, for the metal-poor group the hint for a correlation
is only given by a dozen points (less than 5\% of the total population of
these groups) at rather high metallicity. The metallicity range appears very
narrow for the remaining stars. We then suspect that this group is 
essentially monometallic, the observed spreads in [Fe/H] and [m/H] being only 
due to observational errors, save for very few contaminants, probably assigned
erroneously to this group because of a low measured value of [La/Fe]. In addition,
this plot suggests that for this group there is only a limited contamination by 
AGB stars (which should manifest as objects with low [m/H] for their [Fe/H],
being bluer than RGB stars).
Indeed, some stars scatter in this region of the plot, but they are very few.

As mentioned above, there is an obvious correlation between [Fe/H] and [m/H] for
the intermediate metallicity groups. This suggests that there is a real spread
in metallicity among these groups. However, a closer look reveals that the slope
of the relation between [Fe/H] and [m/H] is significantly smaller than unity,
that is [Fe/H] varies much more than [m/H]. In addition, average [m/H] values for
the different metal intermediate groups (which differs in their [Na/O] value,
that is their location along the Na/O anticorrelation) also differs much less
than what is observed for [Fe/H]. We will come back later to this very interesting
point. We also note that a few points scattered below the bulk of the points in
this plot may be interpreted as AGB contaminants.

Finally, there is some correlation between [Fe/H] and [m/H] also for the metal-rich
group, suggesting that also in this case there is some real spread in metallicity.
Our data are not good enough to conclude if there is a continuous spread, or
rather two or more discrete values (as suggested by Johnson and Pilachowski, 2010). 
More accurate analysis of these stars might establish this interesting point.

\begin{table}[htb]
\centering
\caption[]{Abundances of $\alpha-$elements}
\begin{tabular}{cccccc}
\hline
Group&		&[$\alpha$/Fe]&[Si/Fe]&	[Ca/Fe]& [Ti/Fe]\\
\hline
\multicolumn{6}{c}{Metal-Poor groups}\\
\hline
\#4 &avg	&0.20	& 0.24	& 0.23	& 0.13	\\
	&st.dev.&0.08	& 0.12	& 0.08	& 0.11	\\
\#6 &avg	&0.21	& 0.26	& 0.25	& 0.11	\\
	&st.dev.&0.08	& 0.15	& 0.09	& 0.12	\\
\hline
\multicolumn{6}{c}{Metal-Intermediate groups}\\
\hline
\#5 &avg	&0.27	& 0.32	& 0.30	& 0.20	\\
	&st.dev.&0.12	& 0.14	& 0.13	& 0.18	\\
\#1 &avg	&0.28	& 0.32	& 0.32	& 0.18	\\
	&st.dev.&0.11	& 0.17	& 0.11	& 0.13	\\
\#3 &avg	&0.32	& 0.46	& 0.34	& 0.16	\\
	&st.dev.&0.08	& 0.12	& 0.10	& 0.13	\\
\#2a &avg	&0.36	& 0.44	& 0.36	& 0.28	\\
	&st.dev.&0.08	& 0.16	& 0.10	& 0.16	\\
\hline
\multicolumn{6}{c}{Metal-Rich group}\\
\hline
\#2b &avg	&0.36	& 0.40	& 0.30	& 0.38	\\
	&st.dev.&0.14	& 0.22	& 0.12	& 0.18	\\
\hline
\end{tabular}
\label{t:tab1b}
\end{table}

\subsection{Abundances of $\alpha-$elements}

The results we obtain for the $\alpha-$elements can be summarized as follows (see
Table~\ref{t:tab1b} for the abundances of the individual elements). The
two most metal-poor groups have the same value of [$\alpha$/Fe]=$0.20\pm 0.01$,
within the small statistical error bars. Systematic errors are likely much larger 
than this tiny statistical errors. The [Si/Fe] overabundances (on average, 0.25 dex
for these two groups) are slightly larger than those obtained
for Ca and Ti (0.24 and 0.12 dex), with almost negligible differences between
groups \#4 and \#6. On the whole, we cannot avoid to notice that the $\alpha-$excess is quite
modest, with respect to typical values found in halo stars, and more similar to
those found in dwarf Spheroidals at this metallicity. 

On the other hand, the
intermediate metallicity groups not only provide on average larger $\alpha-$excess,
but [$\alpha$/Fe] seems to be correlated with Na abundances, and
anti-correlated with O ones. Most of this trend is due to the inclusion of Si
among the elements used to estimate $\alpha$/Fe, since there may be some leakage
in the Mg-Al cycle producing Si (see Yong et al. 2005, and Carretta et al. 2009b).
This is clearly present among this group of stars, as also discussed by
Johnson \& Pilachowski (2010). The average values of $\alpha$/Fe for the Fe-intermediate
groups are however larger than simply due to this effect. This may be seen e.g.
by comparing the O-rich group \#5 with
respect to the Fe-poor, O-rich group \#4 ($0.27\pm 0.01$ vs. $0.20\pm 0.01$), or
by comparing the average overabundances of Ca (0.33 vs 0.24) and Ti (0.20 vs 0.12).
This suggests
that the difference between the Fe abundances of the metal-poor and
metal-intermediate groups (only 0.17~dex) is
to be attributed to core collapse and not thermonuclear SNe.

Similarly, the large [$\alpha$/Fe] value obtained for the metal rich group \#2b
may only be explained by core collapse SNe contribution. This result has been already
discussed in Johnson and Pilachowski (2010). For this group, there is a correlation between
the O and $\alpha-$element overabundances.

\begin{center}
\begin{figure}
\includegraphics[width=8.8cm]{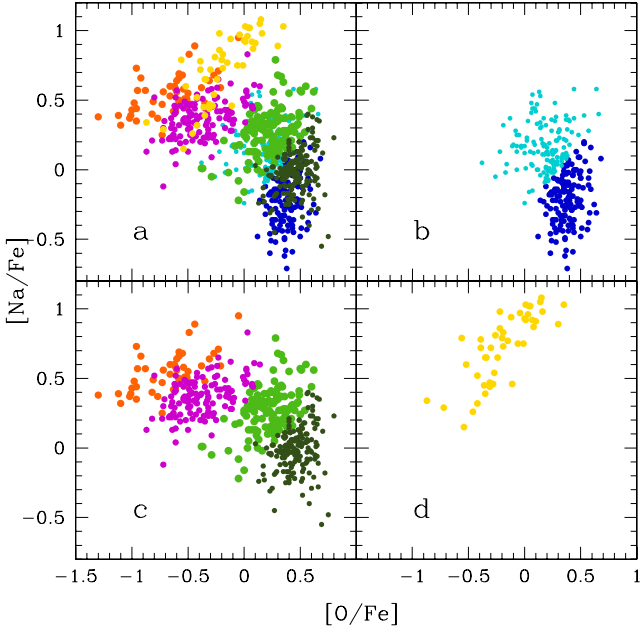}
\caption{Na-O anticorrelation for stars in $\omega$~Cen. Panel a: all stars;
Panel b: metal-poor groups (\#4 and \#6); Panel c: metal-intermediate groups (\#1, \#2a, \#3, and \#5); Panel d:
metal rich group (\#2b). Different colours are for stars 
of the different groups (see Figure~\ref{f:fig1}).}
\label{f:fig5}
\end{figure}
\end{center}

\subsection{Na-O and Al-O anticorrelations in different stellar populations of $\omega$~Cen}

Figure~\ref{f:fig5} shows the Na-O anticorrelation for $\omega$~Cen, with different symbols for
stars attributed to the different groups, as well as the same anticorrelation for
the metal-poor, metal-intermediate, and metal rich groups. As already noticed (see
Carretta et al. 2010a; Johnson \& Pilachowski 2010; Marino et al. 2011) the
run of Na with O is very different in the various metal abundance range. These
differences are even cleaner when using the results of our cluster analysis.
We grouped our data according to the metallicity of the clusters, and plotted
the Na-O anticorrelation separately for metal-poor (\#4 and \#6, blue and cyan, respectively), metal
intermediate (\#1, \#3, \#5 and \#2a, green, violet, black, and red, respectively), and metal rich (\#2b, yellow) groups. The differences in
the Na-O anticorrelation between these three groups are very clear. 

\begin{center}
\begin{figure}
\includegraphics[width=8.8cm]{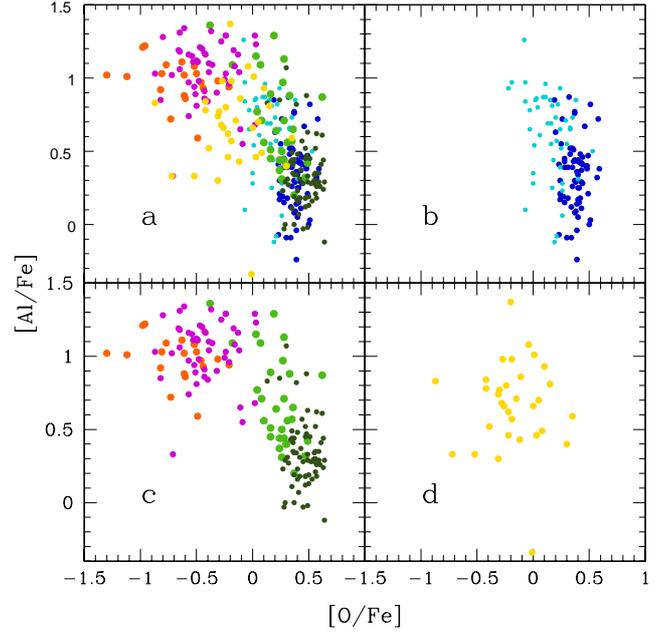}
\caption{Al-O anticorrelation for stars in $\omega$~Cen. Panel a: all stars;
Panel b: metal-poor groups (\#4 and \#6); Panel c: metal-intermediate groups (\#1, \#2a, \#3, and \#5); Panel d:
metal rich group (\#2b). Different colours are for stars 
of the different groups (see Figure~\ref{f:fig1}).}
\label{f:fig6}
\end{figure}
\end{center}

The Na-O anticorrelation is quite modest among the metal-poor groups, which include 
280 stars (35\% of the total): the interquartile of [Na/O] IQR(Na/O) is only 0.53 for this group. 
When compared to those observed in other globular clusters, such a small value is similar 
to that found for 47 Tuc, M3 or M92. Group \#4 (blue) can be identified with the P-star class
and group \#6 (cyan) with the I-star class of Carretta et al. (2009a). If this interpretation
is correct, P stars are the majority of this group ($53\pm 4$\%), a value slightly
larger than that of NGC~2808. Al abundances
are available for 114 stars of these groups (69 in group \#4, and 45 in group \#6).
Al abundances are correlated with those of Na, albeit with some scatter.
Panel a of Figure~\ref{f:fig6} shows the Al-O anticorrelation for these two groups. There
is a clear trend for O-rich stars to be Al-poor, and viz. for O-poor stars. On
average, group \#4 has [Al/Fe]=$0.43\pm 0.04$, while group \#6 has [Al/Fe]=$0.71\pm 0.05$.
However, the star-to-star variations in Al abundances within the metal-poor groups are 
not as large as those observed for Na.

On the other hand, the Na-O anticorrelation is
very extended among the metal intermediate groups, which include 470 stars (59\% of
the total), with IQR(Na/O)=1.10. This value is larger than those of almost all
other globular clusters, the only possible comparison being M54 (Carretta et al. 2010a). If we identify cluster \#5
with the P stars among this group, they make up $34\pm 3$\% of the total. This is distinctly
less than the fraction of P stars within the metal-poor groups. Also, groups \#3 and \#2a
(in total, 176 stars, that is $37\pm 3$\%) can be identified with the E-population. Such
a large fraction of E-stars is not observed in any other globular cluster.
These groups exhibit also a clear Al-O anticorrelation (see panel b of Figure~\ref{f:fig6}), the average [Al/Fe] ratio
steadily increasing between different clusters with decreasing O abundance (note that Al
abundance is available for 32 star of the the most extreme \#2b group, red): 
[Al/Fe]=$0.45\pm 0.04$, $0.84\pm 0.12$, $1.05\pm 0.04$, with 67, 30, and 49 stars for groups \#5 (black), 
\#1 (green), and \#3 (violet),
respectively. The largest scatter for group \#1 agrees well with its intermediate character.
While not extreme, the Al-O anticorrelation is however quite conspicuous over these groups.

Finally, the metal-rich group \#2a (yellow) shows a very clear correlation between Na and O.
This result was already noticed by Johnson \& Pilachowski (2010) and Marino et al. (2011),
but it is even cleaner from our cluster analysis. The Al abundances are available
for only 7 stars, with a possible hint for a correlation with both Na and O abundances
(see panel c of Figure~\ref{f:fig5}). This is clearly different from what observed even in most
metal-rich globular clusters, like NGC6388 and NGC6441, which have metallicities
comparable or even larger than that of this population of $\omega$~Cen (Gratton et al. 2007;
Carretta et al. 2007).

These differences clearly signal a very different nucleosynthesis in the three groups. The
presence of a correlation, rather than anticorrelation, between Na and O in the most metal-rich
group is clearly distinct from what observed in other globular clusters, and indicates
a different class of polluters, possibly related to a prolonged phase of formation. While 
less obvious, the different form of the Na-O and
Al-O anticorrelation seen among the metal-poor and metal-intermediate groups indicates
a less extreme modification and resembles the pair M4-NGC2808 discussed in Carretta et al.
(2009a). It suggests a different range in mass within the class of the polluters
responsible for the observed abundance pattern typically observed in globular clusters.

\begin{center}
\begin{figure}
\includegraphics[width=8.8cm]{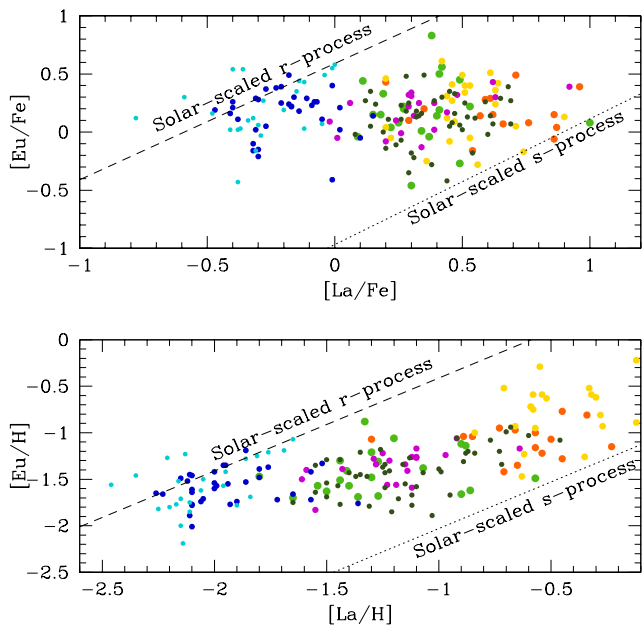}
\caption{Run of [La/Fe] against [Eu/Fe] for stars in $\omega$~Cen. Different colours are for stars 
of the different groups (see Figure~\ref{f:fig1}). Overimposed are the lines corresponding
to a pure solar scaled r- (solid line) and s-contributions (dashed line). } 
\label{f:fig7}
\end{figure}
\end{center}

\subsection{The $n-$capture elements}  

As mentioned in Section 1, $n-$capture elements are one of the basic ingredients 
of the $\omega$~Cen puzzle. In our analysis, we consider two such elements: La, which
in the Sun is mainly produced by the $s-$process, and Eu, which in the Sun is almost
entirely produced by the r-process. La belongs to the second peak of the $s-$process:
very likely, it is produced essentially by the main component, which is thought to be
active in AGB stars of rather low mass ($M<3~M_\odot$). Since these stars have a rather
long lifetime (larger than several $10^8$~yr), the large rise in La abundances observed
between the two most metal-poor groups and the other groups strongly suggest a
corresponding rather large age difference. 

Our cluster analysis only reinforces this concept. The two most metal poor groups, 
essentially distinguished by the [Na/O] value, share the same [La/Fe]=-0.17. Within 
this group there is no evidence for contribution by low mass stars. These two groups 
also have the same Eu abundances ([Eu/Fe]$\sim 0.2$), within the errors. The La/Fe ratio 
is very close to that expected for a solar-scaled r-process (see Figure~\ref{f:fig7}). On the other 
hand, all other groups while having an [Eu/Fe] similar to that observed for the metal-poor
group (again suggesting that heavy elements are enriched by core collapse, and not
thermonuclear SNe), are La-rich. The [La/Fe] values are quite uniform, with a small
increasing trend with metallicity. The comparison of [Eu/La] abundances with solar scaled 
r- and s-contributions clearly indicate that La is produced by some kind of $s-$process.

Hence, either the $s-$process active in $\omega$~Cen is not the main component which
is usually invoked to explain the s-production of elements like La, or there should be
a substantial age difference of several $10^8$~yrs between the formation of the
metal-poor and of the other components of $\omega$~Cen. As discussed in Marino et al. (2011),
it is not easy to avoid the main component as the major $s-$process active in $\omega$~Cen,
so that the first hypothesis lacks at present of any theoretical support.

A way to express this substantial age difference is to think of $\omega$~Cen as made in 
at least two (and probably more) clearly distinct episodes
of star formation, the first one producing a large - but not exceptional - metal-poor
globular cluster, with the typical abundance pattern observed for this class of objects
(no variation in $\alpha-$, Fe-peak, and $n-$capture elements, and with the usual
Na-O and Mg-Al anticorrelations); and at least a second larger one, occurring later,
which produces a much more complex abundance pattern, similar to that observed in
M54. In addition, a third population is present (the RGB-a), with very peculiar
characters, clearly distinct from those typical of globular clusters.

\section{Comparison with other clustering algorithms}

One of the main limitations of cluster analysis is the difficulty to define
the robustness of the group subdivision obtained with one particular algorithm.
For this reason, it is generally useful to compare results obtained with
different cluster algorithms. To test our results we considered two 
of them, available within the R-package: PAM (Partitioning Around Medoids) and
FANNY (Fuzzy Analysis).

PAM is described in chapter 3 of Kaufman and Rousseeuw (1990). Similarly to
k-means, PAM is a partitioning method, but it uses medoids\footnote{Medoids are 
representative objects of a data set or a cluster within a data set whose average 
dissimilarity to all the objects in the cluster is minimal. Medoids are similar 
in concept to means or centroids, but medoids are always members of the data set.} 
, that is real objects, to 
represent the clusters, rather than average values used by k-means. Practically, 
these algorithms select $k$\ 
representative objects arbitrarily. For each pair of non-selected object $h$\
and selected object $i$, the total swapping cost $TC_{ih}$ is calculated.
For each pair of $i$ and $h$, if $TC_{ih} < 0$, $i$\ is replaced by $h$.
Each non-selected object is then assigned to the most similar representative object
and the last two steps are repeated until there is no change. Briefly, PAM starts from 
an initial set of medoids and iteratively replaces one of the medoids by one of
the non-medoids if it improves the total distance of the resulting clustering.
PAM is more robust than 
k-means in the presence of noise and outliers because a medoid is less influenced 
by outliers or other extreme values than a mean. However, our sample contains
no obvious outlier, so this should not be too much a worry in the present case.

FANNY is a fuzzy clustering algorithm. In fuzzy clustering, each point has a degree 
of belonging to clusters, as in fuzzy logic, rather than belonging completely to just 
one cluster (as in crisp logic); thus, points on the edge of a cluster may be in the 
cluster to a lesser degree than points in the center of cluster. The membership of 
observation $i$ to cluster $v$ is denoted by $u(i,v)$. The memberships are non negative, and for a fixed observation $i$,
they sum to 1. The particular method FANNY stems from chapter 4 of Kaufman and Rousseeuw
(1990). Fanny aims to minimize the objective function:
\begin{equation}
\sum\limits_{v=1}^k
\sum\limits_{i,j} u(i,v)^r u(j,v)^r d(i,j)) / (2 \sum\limits_{j} u(j,v)^r)
\end{equation}
where $n$\ is the number of observations, $k$ is the number of clusters, $r$\ is the 
membership exponent and $d(i,j)$\ is the dissimilarity between observations $i$\ and $j$.
Note that $r \rightarrow 1$\ gives increasingly crisper clusterings whereas 
$r \rightarrow \infty$ leads to complete fuzziness. After some trials, we adopted
$r=1.5$\ for the present application, which produces groups of comparable sizes.

Table~\ref{t:tab2} gives the matrices of correspondences between groups defined by k-means, and those
obtained with these two other algorithms. In each analysis, we set the number of groups
at 6. Not surprisingly, there is a very good correspondence between the groups
found by k-means and PAM: in fact, once the groups found by the two algorithms
are properly ordered, the matrix of the correspondences is almost diagonal.
The only significant deviations from a pure diagonal occur because about a third of the
stars of the k-means group \#3 (the moderately O-rich stars of intermediate metallicity)
are combined with the O-rich, metal-rich stars of k-means group \#2 into the PAM group \#4;
and because about a fourth of the stars of k-means group \#6 (metal-poor, 
moderately O-poor stars) are combined with the stars of k-means group \#4 (metal-poor, 
O-rich stars) into PAM group \#6.

The correspondence between k-means and FANNY is also quite satisfactory,
with a fairly diagonal matrix once proper identification of groups is made.
In this case the largest deviations from linearity are that roughly half of
the metal-intermediate, moderately O-poor stars of k-means group \#1 are combined
with the metal-poor, moderately O-poor stars of k-means group \#6,
into the FANNY group \#3. Also, a consistent number of stars of k-means group \#6
are combined with the metal-poor, O-rich stars of k-means group \#4 into
FANNY group \#6; and a similar amount of stars of the same group are combined
with the metal-intermediate, moderately O-poor stars of k-means group \#1,
into FANNY group \#1.

We conclude that comparisons between different clustering methods suggests that
the objective subdivision in groups considered in this paper, while in some case
uncertain for individual stars, is on the whole quite robust. In particular, we 
deem that the agreement between completely different clustering methods (based 
either on hard, e.g. PAM/k-means, or soft partitioning, as FANNY) gives a strong 
support to our analysis. 

\begin{table}[htb]
\centering
\caption[]{Matrices of correspondences between clustering according different algorithms}
\begin{tabular}{ccccccc}
\hline
PAM/k-means & \#1 & \#2 & \#3 & \#4 & \#5 & \#6 \\
\hline
\#1 & 129 &   0 &   0 &   0 &   2 &   6 \\
\#4 &   0 &  95 &  47 &   0 &   0 &   0 \\
\#2 &   2 &   1 &  80 &   0 &   0 &   0 \\
\#6 &   0 &   0 &   0 & 133 &   6 &  34 \\
\#5 &   0 &   0 &   0 &  15 & 154 &   0 \\
\#3 &   1 &   0 &   0 &   0 &   0 &  92 \\
\hline
FANNY/k-means & \#1 & \#2 & \#3 & \#4 & \#5 & \#6 \\
\hline
\#1 &  56 &   0 &   0 &   5 &  24 &  27 \\
\#5 &   0 &  96 &  17 &   0 &   0 &   0 \\
\#4 &   5 &   0 & 105 &   0 &   0 &   0 \\
\#6 &   0 &   0 &   0 & 129 &   0 &  29 \\
\#2 &   8 &   0 &   0 &  14 & 138 &   0 \\
\#3 &  63 &   0 &   5 &   0 &   0 &  76 \\
\hline
\end{tabular}
\label{t:tab2}
\end{table}

\section{Discussion}

\subsection{Comparison between chemical groups and sequences on the colour-magnitude diagram}

We next consider the matching of the different groups identified from our cluster analysis
based on chemical composition to the different sequences
found on the colour-magnitude diagrams (see Pancino et al. 2000; Piotto et al. 2005; 
Bellini et al. 2010). This procedure is much easier for those sequences identified on
the RGB, because we may directly identify our stars in the colour magnitude diagram.
From this first method, we may immediately identify our \#2b group (yellow) with the RGB-a sequence
(Pancino et al. 2000; Sollima et al. 2005a). In our sample, this group includes roughly
6\% of the stars. While this value is similar to those found in other analysis (see e.g. Bellini et al.
2009 and Marino et al. 2011),
it is likely somewhat underestimated when compared to the fraction of e.g. main
sequence stars in the corresponding sequence, 
because we are working with a magnitude limited
sample, and RGB-a stars are typically fainter in visual colours. Recently, Bellini et al.
(2010) were able to follow this branch down to the main sequence. Their data suggests
that this population is He-rich, a fact that could not be derived from RGB alone.

A very interesting discovery in $\omega$~Cen is the presence of a quite well populated
blue main sequence (B-MS), which makes up roughly a quarter of the cluster (Bedin et
al. 2004). Piotto et al. (2005) has found that this B-MS is more metal-rich than
the red main sequence (R-MS: [Fe/H]=$-1.2\pm 0.2$ vs [Fe/H]=$-1.6\pm 0.2$), a fact that can only be explained if it is also
much more He-rich ($Y>0.35$ vs the canonical Big Bang value of $Y\sim 0.25$ for the R-MS),
as originally suspected on the basis of star counts by Norris (2004). While it is likely
that both B-MS and R-MS include several subpopulations (see Bellini et al. 2010), we will
consider them as unique groups for the present discussion. For many reasons, we 
expect that the B-MS is related to stars which are very poor in O (and rich in Na)
(see discussion in Johnson \& Pilachowski, 2010). This
has been beautifully confirmed in the case of NGC~2808 by direct observation of stars in the
different sequences by Bragaglia et al. (2010b). We may then tentatively identify the most 
O-poor groups \#3 (violet) and \#2a (red) as the progeny of the B-MS, and by subtraction, the remaining groups 
should be the progeny of the R-MS. This identification is confirmed by several circumstantial 
facts:
\begin{itemize}
\item On total, we assigned 176 stars to these two groups. This makes up a fraction of 
$22\pm 2$\% of the total, which is very similar to the fraction of main sequence stars in 
the B-MS. It is actually slightly lower, but this can be easily attributed to the different 
areas of the cluster sampled by HST photometry and by the present spectroscopy.
\item Both these two groups belong to the intermediate metallicity group of $\omega$~Cen.
On average, they have [Fe/H]=-1.46, to be compared with a value of [Fe/H]=-1.70 that is
obtained for the groups identified with the R-MS. This difference is similar within the
errors to that obtained for main sequence stars by Piotto et al. (2005).
\item Groups \#2b (yellow) and \#3 (violet) are much more centrally concentrated than the remaining metal-poor
and metal-intermediate groups of $\omega$~Cen. This is analogous to the case of the main
sequence, the B-MS being much more centrally concentrated than the R-MS (Sollima et al.
2007, Bellini et al. 2009).
\item Dupree et al. (2011) showed that the intermediate metallicity, Na/Al-rich (and
thus O-poor) stars had strong He-detections (from the chromospheric line at 10800~\AA)
compared to the Na/Al-poor stars; their sample is small, but it supports our results
\end{itemize}

If we then adopt groups \#3 (violet) and \#2a (red) as the progeny of the B-MS population, it turns out 
that the R-MS should itself include at
least two different metallicity components (a metal-poor and a metal-intermediate one),
and each of them should have an O-rich and a moderately O-poor populations (P and I 
components, following Carretta et al. 2009a nomenclature). We might then expect that a
suitable combination of filters should be able to split the R-MS of $\omega$~Cen
into at least 4 components (and likely more, see below), provided accurate enough
photometry is available. This prediction should be compared with the results
obtained by Bellini et al. (2010). 

On the other hand, the B-MS should be more homogeneous, lacking the metal-poor component.
However, it still likely has a spread both in O-deficiency (and then potentially
in helium), as well as in metallicity (see next subsection).


\subsection{Are groups intrinsically homogeneous?}

{\bf The metal-poor groups.} We obtain a low dispersion in [Fe/H] and [m/H]
for these two groups, and a very similar average value. The narrow range
of both [Fe/H] and [m/H] suggests that residuals about the mean 
are due to random errors of measurement. These two groups
also have very similar abundances of all elements but Na and O, with low
scatter. There is then no reason to think of a real spread in metallicity
among these two groups.

\begin{table*}[htb]
\centering
\caption[]{Determination of $Y$ for star of the intermediate metallicity groups}
\begin{tabular}{ccccccccc}
\hline
Group &N. stars & [m/H]  &   [Fe/H] &   [O/Na] & [Fe/H]-[m/H]      &  Offset & d$Y$   &    $Y$      \\
\hline
\#5     &  162    &-1.562  &  -1.575  &    0.47  & $-0.013\pm 0.021$ &  +0.018 &  0.015 & $0.265\pm 0.016$ \\
\#1     &  127    &-1.642  &  -1.696  &   -0.04  & $-0.054\pm 0.020$ &  -0.023 & -0.019 & $0.231\pm 0.017$ \\
\#5+\#1 &  289    &-1.597  &  -1.628  &          & $-0.031\pm 0.014$ &   0.000 &        &                  \\
\\
\#3     &  132    &-1.600  &  -1.523  &   -0.74  & $+0.077\pm 0.018$ &  +0.108 &  0.088 & $0.338\pm 0.015$ \\
\#2a    &   49    &-1.444  &  -1.308  &   -1.20  & $+0.136\pm 0.034$ &  +0.167 &  0.137 & $0.387\pm 0.028$ \\
\hline
\end{tabular}
\label{t:tab3}
\end{table*}

{\bf Metal intermediate groups.} The dispersion in [Fe/H] and [m/H] is larger for
the four intermediate groups, and the two quantities are correlated,
suggesting a real dispersion in metallicity. Even more interestingly, there is a
possible trend of [Fe/H] values with decreasing [Na/O], a similar but much
less pronounced trend appearing also in [m/H]. These facts might be
interpreted in terms of a variation in the helium content, similarly to what
is observed in NGC2808 (Bragaglia et al. 2010a). To discuss
this point, we first notice that a variation of $Y$\ has two effects on
the metal abundance determinations:
\begin{itemize}
\item the temperature of the RGB rises with $Y$, by some d$T$/d$Y$=500 
(D'Antona et al. 2002). Since d($B-V$)/d$T\sim 0.0008$, we have d$(B-V)$/d$Y$=0.4.
On the other hand d[m/H]/d(B-V)=1.4, hence: dY/d[m/H]$\sim -1.8$.
\item the [Fe/H] value rises because the hydrogen content $X$\ decreases when
the helium content $Y$\ increases, that is $X\sim 1-Y$\ decreases (neglecting
the small metallicity term $Z$). The effect is roughly 
d$Y$/d[Fe/H]$\sim $1.5
\end{itemize}
The offset of $Y$ can then be derived by comparing the offsets in [Fe/H] with [m/H] 
between the different groups. Note that the effect of raising $Y$\ on 
[Fe/H] and [m/H] have opposite signs, so that d$Y$/(d[Fe/H]-d[m/H])$\sim 0.82$.
Let us now consider the four intermediate metallicity groups of $\omega$~Cen
(see Table~\ref{t:tab3}). Column 6 of this table gives the difference 
between the values of [Fe/H] and [m/H] obtained for the different groups.
The difference between groups \#5 (black) and \#1 (green) is small, significant at only about 
1 sigma level (simply considering the dispersion of data for individual stars).
That is: the difference in Y between these two groups is not significant. We 
will then combine these two groups, and assume that they have the same Y=0.25 value 
(the cosmological one). We then estimate the $Y$\ value of groups \#3 (violet) and \#2a (red) using
the relations given above. While this derivation is quite rough, this He excess 
agrees very well with estimates based on the colour of MS stars. We note that 
groups \#3 and \#2a might be put together (with a weighted mean of $Y=0.349\pm 0.013$), 
to correspond to the b-MS of $\omega$~Cen. As discussed in the previous section, 
there are several reasons for this identification. Here we add the most important 
one: they appear to have similar He abundance.  Note that the difference in $Y$\ 
value for the MSs of $\omega$~Cen found in Piotto et al. (2005) might have been 
overestimated, because the impact of a variation of $Y$ on [Fe/H] determinations 
was neglected. Hence the difference in metallicity between the B-MS and R-MS may 
actually be a bit smaller than previously thought (although still within the error 
bars of Piotto et al. 2005).

{\bf Metal-rich group}. Group \#2a displays a quite large spread in metallicity,
and correlated [Fe/H] and [m/H] values (see panel d of Figure~\ref{f:fig4}). This indicates a real spread 
in metallicity. There is some hints that points in panel d of Figure~\ref{f:fig4} are concentrated into
two subgroups, possibly suggesting the existence of two distinct populations,
one with [Fe/H]$\sim -1$, and the other more metal-rich, with [Fe/H]$\sim -0.7$.
This has also been suggested by the metallicity distribution function discussed
by Johnson \& Pilachowski (2010). 
More accurate estimates of [Fe/H] are needed to confirm this hint.

\begin{center}
\begin{figure}
\includegraphics[width=8.8cm]{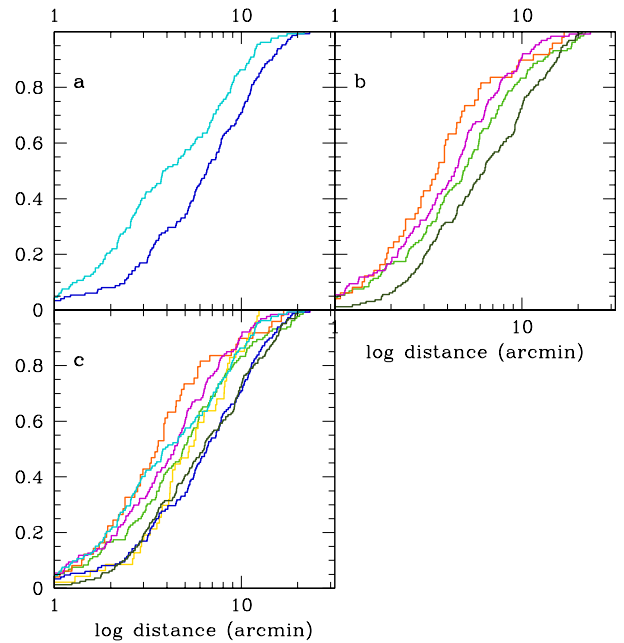}
\caption{Cumulative fraction versus log distance plots. Different colours are for stars 
of the different groups (see Figure~\ref{f:fig1}). Panel a: metal-poor groups only
(\#4 and \#6); panel b: metal-intermediate groups only (\#1, \#2a, \#3, and \#5);
Panel c: all groups. Different colours are for stars of the different groups}
\label{f:fig9}
\end{figure}
\end{center}



\begin{table}[htb]
\centering
\caption[]{Results of Kolmogorov-Smirnov test of consistency between radial distributions of
different groups}
\begin{tabular}{cccccccc}
\hline
   & \#1 & \#2a & \#2b & \#3 & \#4 & \#5 & \#6 \\
\hline
\#1  &       & 0.023 & 0.455 & 0.338 & 0.013 & 0.025 & 0.119 \\
\#2a & 0.023 &       & 0.007 & 0.185 & 0.000 & 0.000 & 0.116 \\
\#2b & 0.455 & 0.007 &       & 0.118 & 0.168 & 0.078 & 0.021 \\
\#3  & 0.338 & 0.185 & 0.118 &       & 0.000 & 0.000 & 0.371 \\
\#4  & 0.013 & 0.000 & 0.168 & 0.000 &       & 0.742 & 0.000 \\
\#5  & 0.025 & 0.000 & 0.078 & 0.000 & 0.742 &       & 0.008 \\
\#6  & 0.119 & 0.116 & 0.021 & 0.371 & 0.000 & 0.008 &       \\
\hline
\end{tabular}
\label{t:tab4}
\end{table}

\subsection{Concentration and kinematics of the groups}

Among the various properties of the groups identified by our cluster
analysis, it is interesting to study the radial density distributions
of the various populations. $\omega$~Cen has a half-light two-body
relaxation time comparable to its age ($\sim 12$~Gyr: see Harris 2010). The
relative distribution of stars reflect then the initial conditions.

Panel c of Figure~\ref{f:fig9} shows the cumulative fraction versus log distance plots.
Examining this figure, we find that the O-rich groups \#4 (blue) and \#5 (black) exhibit 
a nearly identical distribution; all other groups appear more centrally 
concentrated. This makes sense since the other groups are more O-poor than
groups \#4 and \#5, and there appears to be a general correlation between
central concentration and O-deficiency among globular clusters (see
Norris \& Freeman 1979, Kravtsov et al. 2010, Carretta et al. 2010c, 
Lardo et al. 2011, Nataf et al. 2011); the larger central concentration
of the He-rich populations in $\omega$~Cen was already noticed by
Sollima et al. (2007) and Bellini et al. (2009).
It appears also that groups \#1 (green) and \#3 (violet) are rather similar 
in their spatial distribution, while group \#2a (red) 
is clearly more centrally concentrated than the 
other populations; this is not surprising since these are the most O-poor.
stars. All these results are confirmed by Kolmogorov-Smirnov tests (see
Table~\ref{t:tab4}). We interpret this result as an indication that the
second generation stars (I component) of the metal-poor group are more
centrally concentrated than the first generation (P component) stars.
This is in agreement with the radial distribution of the P and I components
found in many monometallic globular clusters by Carretta et al. (2009a).

Panel a of Figure~\ref{f:fig9} shows the metal-poor only plot. There is a very
significant difference between the radial distributions of groups \#4 (blue) and \#6 (cyan),
as confirmed by the Kolmogorov-Smirnov test which yields a probability of 
$1.7\times 10^{-4}$\ that the two groups are extracted from parent populations
having the same radial distributions.

Panel b of Figure~\ref{f:fig9} shows the same plot, but this time for the intermediate metallicity 
groups only. It appears that groups \#1 (green) and \#3 (violet) are rather similar in their spatial
distribution, while group \#2a (O-poor, red) is clearly more centrally concentrated
than the other populations, the opposite holdings for group \#5 (O-rich, black).
Clearly there is a correlation between the degree of O-depletion and
radial concentration; the O-rich stars are more evenly distributed and 
the most O-poor stars are the most centrally concentrated; the "gradient" 
is wonderfully illustrated in this plot.
The Kolmogorov-Smirnov tests of Table~\ref{t:tab4} confirm the significance
of these findings.

\begin{center}
\begin{figure}
\includegraphics[width=8.8cm]{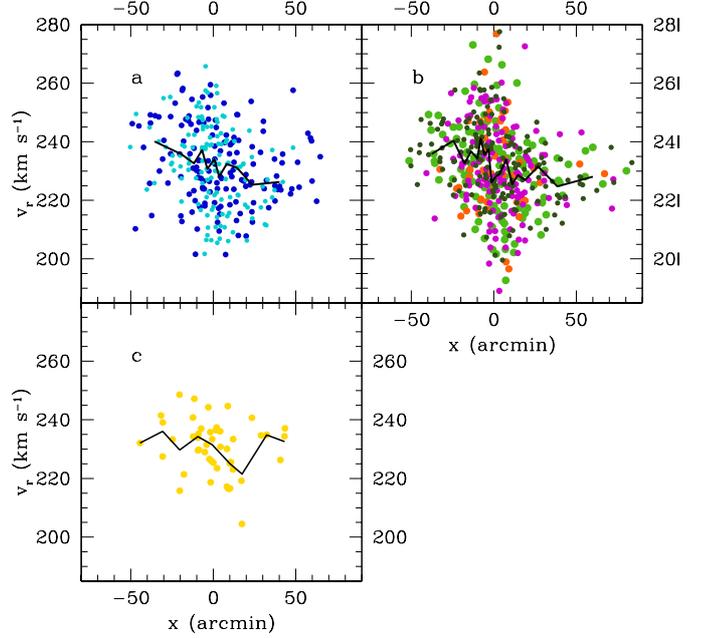}
\caption{Radial velocity vs position along the major axis of $\omega$~Cen. 
Panel a: metal-poor groups
(\#4 and \#6); Panel b: metal-intermediate groups (\#1, \#2a, \#3, and \#5); Panel c:
metal rich group (\#2b). Different colours are for stars 
of the different groups (see Figure~\ref{f:fig1}).}
\label{f:fig12}
\end{figure}
\end{center}

For what concerns kinematics, it is well known that $\omega$~Cen has
a quite conspicuous rotation (Merrit et al. 1997; Norris et al. 1997; 
Sollima et al. 2005b; Reijns et al. 2006; Pancino et al. 2007), the rotation axis appearing to be
close to the minor axis projected on sky. Radial velocities are available
for all stars in our sample, either from Reijns et al. (2006) or from
measurements on the spectra gathered by Johnson \& Pilachowski (2010).
Average values and standard deviations about the mean for the individual
groups of our analysis are listed in the last column of Table~\ref{t:tab1}.
Figure~\ref{f:fig12} shows the rotation
curves we obtain for the various populations. There
is not a large difference between the rotation curves for the various populations,
although data are not adequate for the \#2b group (yellow). However, this last group is
peculiar in being apparently kinematically cooler than the other components,
with a r.m.s. spread of radial velocities of only 9.1 km/s (while values
for the other components are in the range 13-15 km/s: see Table~\ref{t:tab1}).

\section{Conclusions}

$\omega$~Cen is a complex cluster composed of several populations. These
populations differ in many different characteristics (helium, metallicity,
abundances of light elements, central concentration, etc.), and likely
originated in different episodes of star formation, although it is
also possible that some of the observed differences might simply be due to poor
mixing in very extended star forming regions. Disentangling different
populations is a first step in trying to reconstruct this puzzle.

Trying to put this on objective basis, we performed a classical cluster 
analysis on the extensive spectroscopic data recently obtained by Johnson 
and Pilachowski (2010). We used the popular k-means algorithm, but we also 
found similar results with other algorithms which use different approaches, 
so that we deem the main conclusions rather robust.

Our group analysis suggests that stars in $\omega$~Cen might be divided
into three main groups, which likely have a different history: 
\begin{itemize}
\item A metal-poor group, which includes about a third of the total. This is
the simplest population, appearing to be itself quite homogeneous for what concerns
the abundances of most elements, and looks quite similar to a typical globular
cluster. It shows a moderate O-Na anticorrelation, and similarly to other cases,
the O-poor second generation stars are more centrally concentrated
than the O-rich first generation ones. This whole population is La-poor, with a 
pattern of abundances for $n-$capture elements which is very close to a scaled r-process 
one.  
\item A metal-intermediate group, which includes the majority of the cluster stars.
This is a much more complex population, with an internal spread in the abundances
of most elements. It shows an extreme O-Na anticorrelation, with a very numerous
population of extremely O-poor and He-rich second generation stars. This second generation is very
centrally concentrated. The whole population is La-rich, with a pattern 
of the abundances of $n-$capture elements that shows a strong contribution by the $s-$process.
We tentatively suggest that this population is due to a very large episode of
star formation which occurred several hundred million years later than the episode
responsible for the metal-poor group, within the same dwarf galaxy. The composition
difference between the metal-poor and the metal intermediate populations might be
attributed to the chemical evolution within this galaxy, where most (but not all)
of the products of core collapse SNe of the earlier populations were lost, while
virtually all products of the intermediate mass stars that have evolved in this lapse were 
retained. The spread in metallicity within this metal-intermediate population is not very large, and 
we might attribute it either to non uniformities of an originally very 
extended star forming region, or to some ability to retain a fraction of the ejecta 
of the core collapse SNe that exploded first, or both. The first hypothesis 
is favoured by the existence of a correlation between $s-$process and Fe-peak element
abundances, that is more easily explained if primordial. This second episode might have been
coincident in space with the earlier one, as seems to happen in nuclear
star clusters (see e.g. B\"oker 2008). Alternatively, we might think that the metal-poor and metal-intermediate
populations formed as separate clusters within a single dwarf galaxy, and have later
merged after migration toward the center of this galaxy caused by dynamical
friction (see Bellazzini et al. 2008; Johnson \& Pilachowski 2010; Agarwal \& Milosavljevi 2011). 
\item A metal-rich group. This is perhaps the most mysterious group. The presence
of a Na-O correlation, rather than anticorrelation, clearly separates this population
from globular clusters. In particular, we notice that a Na-O anticorrelation is 
present even in most metal-rich globular clusters (Gratton et al. 2007; Carretta et al. 2007), 
indicating that the high metallicity 
is not the reason of the difference between the composition of this group and
that typical of globular clusters. Rather, the range of mass of the stars responsible 
for the metal-enrichment should be different. We suggest that stars over a much wider 
range of masses than typical for globular clusters must be considered: there is
evidence for the contribution of both massive stars ending their life as core-collapse
SNe, and intermediate/small mass stars, producing n-capture elements through the s-process. On the other
hand, there is no evidence for a contribution by SN Ia. While, following Carretta et al. (2010a), we were 
tempted to interpret this group as stars of the host galaxy captured by the cluster 
(like the "nuclear" population seen in M54, which shares the composition of the 
Sagittarius dwarf galaxy), its composition is very different from that typically 
observed in dwarf Spheroidals. In particular, it  shows the typical excess of 
$\alpha-$elements seen in globular clusters, while the metal-rich stars in dwarf Spheroidals
invariably have a relative deficiency of $\alpha-$elements with respect to Fe. 
Furthermore, this population is more centrally concentrated than some of the other
components, and much more kinematically cool, which is 
unexpected for a captured population and looks more consistent with further episodes
of star formation within a nuclear cluster. Incidentally, we note that the lack of a sizeable 
"dwarf spheroidal"-like population in $\omega$~Cen is itself a fact that must be 
explained by scenarios of formation for this intriguing cluster.
\end{itemize}

\begin{acknowledgements}
This research has been funded by PRIN MIUR 20075TP5K9, and by PRIN INAF
"Formation and Early Evolution of Massive Star Clusters". This material uses work 
supported by the National Science Foundation under award No. AST-1003201 to CIJ. 
CAP gratefully acknowledges support from the Daniel Kirkwood Research Fund at Indiana 
University.
\end{acknowledgements}

\end{document}